\documentclass[a4paper,12pt]{article}
\usepackage[dvips]{graphicx}
\title{Propagation of the main signal in a dispersive Lorentz medium}
\author{A.~Ciarkowski\\
\normalsize Institute of Fundamental Technological Research\\
\normalsize Polish Academy of Sciences}
\date{}

\def\o{\omega}
\def\w{\tilde}
\def\t{\theta}
\def\b{{\cal B}}
\def\g{\gamma}
\def\r{\rho}
\def\os{\o_{s}}
\def\l{\lambda}
\def\d{\delta}
\begin{document}

\maketitle

\begin{abstract}
Evolution of the main signal in a Lorentz dispersive medium is
considered. The signal propagating in the medium is excited by a
sine-modulated pulse signal, with its envelope described by a
hyperbolic tangent function. Both uniform and non-uniform
asymptotic representations for the signal are found. It is shown
when the uniform representation can be reduced to the non-uniform
one. The results obtained are illustrated with a numerical
example.
\end{abstract}

\section{Introduction}
Investigations on propagation of pulse signals in dispersive media
date back to the beginning of 20th century. The fundamental
research in this area is due to Sommerfeld \cite{so;14} and
Brillouin \cite{br;14,br;60}. Although steady interest in this
kind of propagation was observed in the literature since then, a
new impetus has been added recently due to new applications of the
theory in fiber-optics communication and integrated-optics. Also,
the knowledge of pulse propagation in a dispersive medium, and of
accompanying electromagnetic energy losses in the medium, became
of vital importance in radiotherapy. A significant contribution to
the research on dispersion phenomena in Lorentz media is due to
Oughstun and Sherman \cite{ou;97}. Equipped with better
computation techniques and advanced asymptotic methods, they
extended the analysis to models more closely reflecting practical
applications. In particular, they considered signals with a finite
rise time and employed uniform asymptotic expansions in their
analysis. (Uniform expansions remain valid as their parameters
vary while non-uniform expansions break down at some parameters
values.)

Here, we also consider the evolution of the main signal excited in
a Lorentz dispersive medium by a signal with a finite rise time.
However, unlike Oughstun and Sherman work, where the envelope of
the initial signal is described by an everywhere smooth function
of time which tends to zero as time goes to minus infinity, our
exciting signal is switched abruptly at a finite time instant, and
vanishes identically for earlier times. In the analysis carried
out in this paper we apply the Bleistein and Handelsman
\cite{bl;75} theory of uniform asymptotic evaluation of integrals
with nearby saddle point and an algebraic singularity. We show,
how the uniform representation of the evolution of the main signal
reduces to the non-uniform representation, which can otherwise be
obtained by residues.

The results obtained here are illustrated with a numerical
example.

\section{Formulation of the problem}
Consider the problem of an electromagnetic plane wave propagation
in a homogeneous, isotropic medium, whose dispersive properties
are described by the Lorentz model of resonance polarization. The
complex index of refraction in the medium is given by \cite{ou;97}
\begin{equation}\label{e1}
n(\o) = \left( 1 - \frac{b^2}{\o^2 - \o_0^2 + 2 i \delta \o}
\right)^{1/2}.
\end{equation}
Here, $b^2=4\pi N e^2/m$, where $N$, $e$ and $m$ represent the
number of electrons per unit volume, electron charge and its mass,
respectively, $\delta$ is a damping constant and $\o_0$ is a
characteristic frequency.

Let the signal $A_0(t)$ in the plane $z=0$ be a sine wave of a
fixed real frequency $\o_c$ with its envelope described by a real
function $u(t)$, identically vanishing for $t<0$, i.e.
\begin{equation}\label{e2}
A_0(t)=\left\{
\begin{array}{ll}
0 &               t<0 \\
u(t)\sin(\o_ct) & t\ge 0.
\end{array}
\right.
\end{equation}
Then arbitrary component of the wave propagating in the direction
of increasing $z$ (or of a corresponding Hertz vector) can be
represented in the medium by the scalar function \cite{ou;97}
\begin{equation}\label{e3}
A(z,t)=\frac{1}{2\pi}\,\mbox{Re}\left\{i \int_{i a-\infty}^{i a+\infty}
\w{u}(\o-\o_c)\exp \left[\frac{z}{c}\phi(\o,\t)\right]\,d\o\right\},
\end{equation}
where $\w{u}(\o)$ is the Laplace transform of $u(t)$. The complex phase
function $\phi(\o,\t)$ is given by
\begin{equation}\label{e4}
\phi(\o,\t)=i\o[n(\o)-\t],
\end{equation}
where the dimensionless parameter
\begin{equation}\label{e5}
\t=\frac{c t}{z}
\end{equation}
describes the space-time point $(z,\,t)$.

It is here assumed that the envelope of the incident pulse is
described by
\begin{equation}\label{e6}
u_\beta(t)=\left\{
\begin{array}{ll}
0 &               t<0 \\
\tanh \beta t   &  t\ge 0,
\end{array}
\right.
\end{equation}
where the parameter $\beta\ge 0$ determines the rate of the pulse
growth.

\noindent The Laplace transform of $u(t)$ is
\begin{equation}\label{e7}
\w{u}_\beta(\o)=
\frac{1}{\beta}\b\left(-\frac{i\o}{2\beta}\right)-\frac{i}{\o},
\hskip.5in \mbox{Im}\;\o>0,
\end{equation}
and the beta function $\b$ is related to the psi function $\psi$
by \cite{rg;51}
\begin{equation}\label{e8}
\b(x) = \frac{1}{2}\left[\psi\left(\frac{x+1}{2}\right) -
\psi\left(\frac{x}{2}\right)\right].
\end{equation}
By using (\ref{e7}) in (\ref{e3}) we obtain the formula
\begin{equation}\label{e9}
A(z,t)=\frac{1}{2\pi}\,\mbox{Re}\left\{i \int_{i a-\infty}^{i
a+\infty} \w{u}_\beta(\o-\o_c)
e^{\frac{z}{c}\phi(\o,\t)}\,d\o\right\},
\end{equation}
which describes the dynamics of the signal excited at $z=0$ by
$A_0(t)$, and propagating in the Lorentz dispersive medium in the
direction of growing $z$. The uniqueness of this solution is
proved in Sec.~2 of \cite{br;60} .

In this work we study the poles contribution to the asymptotic
expansion of $A(z,t)$. We denote this contribution by $A_c(z,t)$
and find both non-uniform and uniform asymptotic expressions for
it.

\section{Non-uniform asymptotic expression for\newline $A_c(z,t)$}
In finding an asymptotic expansion of the integral defined by
(\ref{e9}) it is essential to determine the location of its
critical points, including saddle points and the poles in the
complex $\o$-plane. The equation governing the location of the
saddle points does not seem to be solvable exactly. Instead,
different approximate solutions were obtained by Brillouin
\cite{br;60}, Kelbert and Sazonov \cite{ks;96}, and Oughstun and
Sherman \cite{ou;97} to describe the location. Recently, a new
approximation for this location was obtained in \cite{ac;98}. In
this work, however, we shall employ a numerical approximation of
the saddle point solution obtained with the help of the
\emph{Mathematica} computer program, and based on interpolation
techniques.

\noindent As in Oughstun and Sherman study, we deform the original
contour of integration to the Olver type contour $P(\t)$
\cite{ol;70} which passes through the near and distant saddle
points. The pole contribution to the asymptotic expansion of
(\ref{e9}) occurs if in the process of contour deformation one or
more poles of $\w{u}_\beta(\o)$ are crossed. From the series
representation of the function $\w{u}_\beta(\o)$ \cite{ac;97}
\begin{equation}\label{e10}
\w{u}_\beta(\o)=\frac{i}{\o}-2i\left(\frac{1}{\o+2i\beta}-
\frac{1}{\o+4i\beta}+\cdots\right),
\end{equation}
it follows that the integrand in (\ref{e9}) has an infinite set of
poles $\o=\o_c-2im\beta$, $m=0,1,2,\cdots$, in the half-plane Im
$\o\le 0$, which are located along a line, parallel to the $\o$
imaginary axis. If $\beta$ is big enough, only the real pole
$\o=\o_c$ is of importance, since the remaining poles are not
crossed in the process of contour deformation. If, however,
$\beta$ is small, one or more of the remaining poles can be
crossed, and their contributions must then be taken into account.

Let $\t_s$ be the value of $\t$, at which the deformed contour
crosses the pole at $\o=\o_c$ in (\ref{e9}), $\o_c$ being real and
positive. Then, by the Cauchy theorem,
\begin{equation}\label{e11}
A_c(z,t)=\left\{
\begin{array}{ll}
\displaystyle
0, & \t<\t_s, \\[1.5ex]
e^{-\frac{z}{c}\o_c n_i(\o_c)} \sin{[\frac{z}{c}\o_c(n_r(\o_c)-\t)]}, & \t>\t_s.
\end{array}
\right.
\end{equation}
Here, $n_r(\o_c)$ and $n_i(\o_c)$ stand for real and imaginary
parts of $n(\o_c)$, respectively.

Upon introducing the amplitude attenuation coefficient \cite{ou;97}
\begin{equation}\label{e12}
\alpha(\o_c)=\frac{\o_c}{c} n_i(\o_c),
\end{equation}
and the propagation factor
\begin{equation}\label{e13}
\zeta(\o_c)=\frac{\o_c}{c} n_r(\o_c),
\end{equation}
$A_c(z,t)$ can be written down as
\begin{equation}\label{e14}
A_c(z,t)=\left\{
\begin{array}{ll}
\displaystyle
0, & \t<\t_s, \\[1.5ex]
e^{-z\alpha(\o_c)} \sin{[\zeta(\o_c)z-\o_c t]}, & \t>\t_s.
\end{array}
\right.
\end{equation}
It then follows that for real and positive $\o_c$ the pole
contribution to the asymptotic expansion of $A(z,t)$ oscillates in
time at the frequency $\o_c$ and decreases along its propagation
distance $z$ with time independent attenuation coefficient
$\alpha(\o_c)$.

The pole contribution (\ref{e14}) represents a discontinuous
function of $\t$, while the integral representation of $A(z,t)$
changes continuously with $\t$. As pointed in \cite{ou;97}, this
fact is of little significance if $z$ is finite and the pole is
bounded away from the dominant saddle point at $\o=\os$. Denote
$X(\o,\t)=\mbox{Re}\;\phi(\o,\t)$. Then $e^{-(z/c) X(\o_c,\t)}$ is
negligible in comparison to the saddle point contribution which
has the magnitude $e^{-(z/c) X(\os,\t)}$. Hence, the discontinuous
behaviour of $A_c(z,t)$ is then also negligible.

\section{Uniform asymptotic expression for $A_c(z,t)$}
The situation becomes different if the dominant saddle point
approaches the pole at $\o=\o_c$. In this case $X(\o_c,\t)$ is
comparable with $X(\os,\t)$ and so are the pole and the branch
point contributions to the asymptotic expansion of $A(z,t)$. To
obtain a continuous asymptotic representation for $A_c(z,t)$, a
uniform approach, as proposed by Bleistein and Handelsman
\cite{bl;75} will here be used.

Let us consider the first pole at $\o=\o_c$. From (\ref{e9}) and
(\ref{e10}) it follows that the function to be asymptotically
evaluated is
\begin{equation}\label{e15}
A_c(z,t)=-\frac{1}{2\pi}\,\mbox{Re}\left\{\int_{P(\t)}
\frac{e^{\l\phi(\o,\t)}}{\o-\o_c} \,d\o\right\},
\end{equation}
where $\l=z/c$. The function $\phi(\o,\t)$ has a saddle point at
$\o=\os$, where $\os=\os(\t)$. (One can verify that the contour
$P(\t)$ through a near (distant) saddle point $\o=\os$ makes the
angle $\pi/4$ ($3\pi/4$) with the real axis.)

In accordance with the  Bleistein and Handelsman method we
introduce a new variable of integration $\tau$, defined by
\begin{equation}\label{e17}
\phi(\o,\t)=-\frac{\tau^2}{2}-\g\tau+\r=\Psi(\tau,\t).
\end{equation}
The quantities $\g$ i $\r$ are chosen so that $\tau=-\g$ is the
image of the saddle point $\o=\os$ and $\tau=0$ is the image of
$\o=\o_c$. Then,
\begin{equation}\label{e18}
\r(\t)=\phi(\o_c,\t) \hbox{\ \ \ and \ \ \
}\g(\t)=\sqrt{2[\phi(\os,\t)-\phi(\o_c,\t)]}.
\end{equation}
The complex-valued function $\g(\t)$ is defined such that it is a
smooth function of $\t$ when its argument varies in the interval
$-\pi<$ Arg $\g\le\pi$.

One finds from (\ref{e17}) that
\begin{equation}\label{e19}
\tau+\g=\sqrt{2[\phi(\os,\t)-\phi(\o,\t)]},
\end{equation}
and hence for $\o$ near $\os$:
\begin{equation}\label{e20}
\tau+\g\approx\sqrt{-\phi_{\o\o}(\os,\t)}(\o-\os)[1+O(\o-\os)].
\end{equation}
The steepest descent path through the saddle point $\tau=-\g$ runs
parallel to the real axis.
Upon using (\ref{e17}) in (\ref{e15}) the function $A_c(z,t)$ takes the
form
\begin{equation}\label{e21}
A_c(z,t)=-\frac{1}{2\pi}\,\mbox{Re}\left\{\int_{C(\t)}
\frac{G_0(\tau,\t)}{\tau}e^{\l\Psi(\tau,\t)} \,d\tau\right\},
\end{equation}
where
\begin{equation}\label{e22}
G_0(\tau,\t)=\frac{\tau}{\o-\o_c}\frac{d\o}{d\tau}
\end{equation}
and $C(\t)$ is the image of $P(\t)$ under (\ref{e17}).

We now expand $G_0$ in the form
\begin{equation}\label{e23}
G_0(\tau,\t)=a_0+a_1\tau+\tau(\tau+\g)H_0(\tau,\t),
\end{equation}
where $H_0(\tau,\t)$ is a regular function of $\tau$. Since the
last term vanishes at both critical points $\tau=-\g$ and
$\tau=0$, the coefficients $a_0$ and $a_1$ are given by
\begin{equation}\label{e24}
a_0=G_0(0,\t)\hbox{\ \ \ \ and \ \ \ }a_1=\frac{G_0(0,\t)-G_0(-\g,\t)}{\g}.
\end{equation}
By L'Hospital's rule:
\begin{equation}\label{e25}
\lim_{\tau\rightarrow
0}{\frac{\o-\o_c}{\tau}}=\lim_{\tau\rightarrow
0}{\frac{d\o}{d\tau}},
\end{equation}
and hence
\begin{equation}\label{e26}
G_0(0,\t)=1.
\end{equation}
Furthermore, from (\ref{e20})
\begin{equation}\label{e27}
\lim_{\tau\rightarrow
-\g}{\frac{d\o}{d\tau}}=-\frac{1}{\sqrt{-\phi_{\o\o}(\os,\t)}},
\end{equation}
and thus,
\begin{equation}\label{e28}
G_0(-\g,\t)=
-\frac{\g}{\os-\o_c}\;\frac{1}{\sqrt{-\phi_{\o\o}(\os,\t)}}.
\end{equation}
In this manner we obtain
\begin{equation}\label{e29}
a_0=1 \hbox{\ \ \ \ and \ \ \ }
a_1=\frac{1}{\g}+\frac{1}{\os-\o_c}\;\frac{1}{\sqrt{-\phi_{\o\o}(\os,\t)}}.
\end{equation}
If now (\ref{e23}) is inserted into (\ref{e21}), and the resulting
canonical integrals (\cite{bl;75}) are expressed by special functions,
the following result is found
\begin{equation}\label{e30}
A_c(z,t)=\frac{1}{2\pi}\,\mbox{Re}\left\{e^{\l\r}\left[
W_{-1}(\sqrt{\l}\g)+\frac{a_1}{\sqrt{\l}}W_0(\sqrt{\l}\g)
\right]+R_0(\l,\t)\right\},
\end{equation}
where,
\begin{equation}\label{e31}
W_0(z)=\sqrt{2\pi}e^{\frac{z^2}{2}}\hbox{\ \ \ \ and \ \ \ }
W_{-1}(z)=i\int_{-iz}^\infty e^{-\frac{s^2}{2}}ds.
\end{equation}
The remainder of the expansion, $R_0$, is given by
\begin{equation}\label{e32}
R_0(\l,\t)=\l^{-1}\int_{C(\t)} G_1(\tau,\t)e^{\l\Psi(\tau,\t)}
\,d\tau,
\end{equation}
with
\begin{equation}\label{e33}
G_1(\tau,\t)=\tau\frac{dH_0}{d\tau}.
\end{equation}
In arriving at (\ref{e32}) we integrated the last term in (\ref{e23}) by
parts and neglected the boundary contributions as being asymptotically
negligible.

The function $W_{-1}$ can be expressed in terms of the
complementary error function
$\mbox{erfc}(z)=2/(\sqrt{\pi})\int_z^\infty e^{-s^2}ds$. By using
(\ref{e29}) and (\ref{e31}) in (\ref{e30}) we arrive at the
following uniform asymptotic representation
\begin{eqnarray}\label{e34}
\lefteqn{A_c(z,t)\sim}  \\
& & \hspace{-2pt}\mbox{Re}\left\{ e^{\l\r}\left[ \frac{i}{2}\;
\mbox{erfc}\left(i\g\sqrt{\frac{\l}{2}}\right)-
\frac{e^{\frac{\l\g^2}{2}}}{\sqrt{2\pi\l}} \left(
\frac{1}{\g}+\frac{1}{(\os-\o_c)\sqrt{-\phi_{\o\o}(\os,\t)}}
\right) \right]\right\},\nonumber
\end{eqnarray}
of the main signal in the medium, provided only the first pole
$\o=\o_c$ interacts with the saddle point.

This asymptotic formula applies for any $\sqrt{\l}\g$. In
particular, if $\g\rightarrow 0$, the components of the last
parentheses blow up, but their sum remains bounded.

If $\sqrt{\l}|\g|$ is large, the error function in (\ref{e34}) can
be approximated by its asymptotic expansion (comp.\ \cite{ac;99})
\begin{equation}\label{e35}
\mbox{erfc}(iy)=\eta(y)-e^{y^2}\left[\frac{i}{\sqrt{\pi}y}
+O(y^{-3})\right],
\end{equation}
where
\begin{equation}\label{e36}
\eta(y)=\left\{
\begin{array}{ll}
0, & -\pi<\mbox{Arg}(y)<0,\\[1ex]
1, & \mbox{Arg}(y)=-\pi \mbox{\ \ or\ \ } 0,\\[1ex]
2, & 0<\mbox{Arg}(y)<\pi.
\end{array}
\right.
\end{equation}
Upon using this expansion in (\ref{e34}), the non-uniform
asymptotic representation of the main signal evolution results:
\begin{equation}\label{e37}
A_c(z,t)\sim\mbox{Re}\left\{ \frac{ie^{\l\r}}{2}\eta(\g)+
\frac{e^{\l
\phi(\os,\t)}}{\os\sqrt{-\phi_{\o\o}(\os,\t)}}\right\}.
\end{equation}
If $\mbox{Arg}(\g)<0$, which occurs when the pole at $\o=\o_c$ is
located to the right with respect to the contour $P(\t)$, the main
signal is absent in $A_c(z,t)$, and only the term that appears is
that proportional to $(-\phi_{\o\o}(\os,\t))^{-1/2}$. This term
can be interpreted as due to the saddle point $\os$.

If, $\mbox{Arg}(\g)>0$, which occurs after the contour crosses the
pole, then in addition to the term proportional to
$(-\phi_{\o\o}(\os,\t))^{-1/2}$, a new term appears
\begin{equation}\label{e38}
\mbox{Re}\left\{ie^{\l\r}\right\}=e^{-\frac{z}{c}\o_c n_i(\o_c)}
\sin\left[\frac{z}{c}\o_c[n_r(\o_c)-\t]\right].
\end{equation}
It represents the main signal and its form fully agrees with
(\ref{e11}).

In this manner we have obtained both uniform and non-uniform
asymptotic representations for the evolution of the main signal in
the medium, which are described by (\ref{e34}) and (\ref{e37}),
respectively. While the uniform representation applies for any
value of $\sqrt{\l}\g$, the non-uniform representation is valid
only for sufficiently large values of $\sqrt{\l}|\g|$.

One remark should now be made. The applied theory assumes that the
saddle point is of the first order, i.e.\ $\phi_{\o\o}(\os,\t)$ is
never zero. In the present context this assumption is satisfied
everywhere except for the special value of $\t=\t_1$, where two
coalescing near simple saddle points merge on the $\o$-imaginary
axis to form a saddle point of the second order. Hence
$\phi_{\o\o}=0$ at $\t=\t_1$, and consequently both asymptotic
representations of $A_c(z,t)$, as given by (\ref{e34}) and
(\ref{e37}), are there invalid. Therefore, strictly speaking, if
the carrier frequency $\o_c$ lies below anomalous dispersion
region, (\ref{e34}) is a uniform if $\t>\t_1$.

\section{Numerical example}
A numerical example is now given to illustrate the results
obtained in the previous section. It is assumed that the Lorentz
medium is described by Brillouin's choice of medium parameters
\begin{equation}\label{e39}
b=\sqrt{20.0}\times 10^{16} s^{-1}, \quad \o_0=4.0\times 10^{16}
s^{-1}, \quad \d=0.28\times 10^{16} s^{-1},
\end{equation}
and additionally, $\l=3.0\times 10^{-15}$, and $\o_c=2.0\times
10^{16}s^{-1}$. The latter choice implies that in this example the
saddle point in question is the near one.

Let us first suppose that the parameter $\beta$ in (\ref{e6}) is
large enough, say of the order of $10^{17}$ or more, to ensure
that the second pole $\o_{c2}=\o_c-2i\beta$ is sufficiently
distant from the contour $P$, and, in particular, it is not
crossed in the process of the contour deformation. Then only the
real pole at $\o=\o_c$ is of interest. Under this assumption the
real and imaginary parts of $\g(\t)$, as given by (\ref{e17}), are
shown in Fig.~1. In order to determine numerical values of the
function $\o_s(\t)$ an interpolation technique provided by the
\emph{Mathematica} computer program has been employed. The
evolution of the main signal, as predicted by the uniform
asymptotic representation (\ref{e34}), is depicted in Fig.~2. The
anomaly in the plot at $\t=\t_1\approx 1.5$, results from
vanishing $\phi_{\o\o}(\os,\t)$ at $\t=\t_1$, and, as discussed in
the previous section, the result obtained from (\ref{e34}) breaks
down there. Fig.~3 shows the corresponding plot obtained from
(\ref{e34}), in which the term proportional to
$1/\sqrt{-\phi_{\o\o}}$, has been dropped.

\begin{figure}
\begin{center}
\includegraphics[width=.7\textwidth]{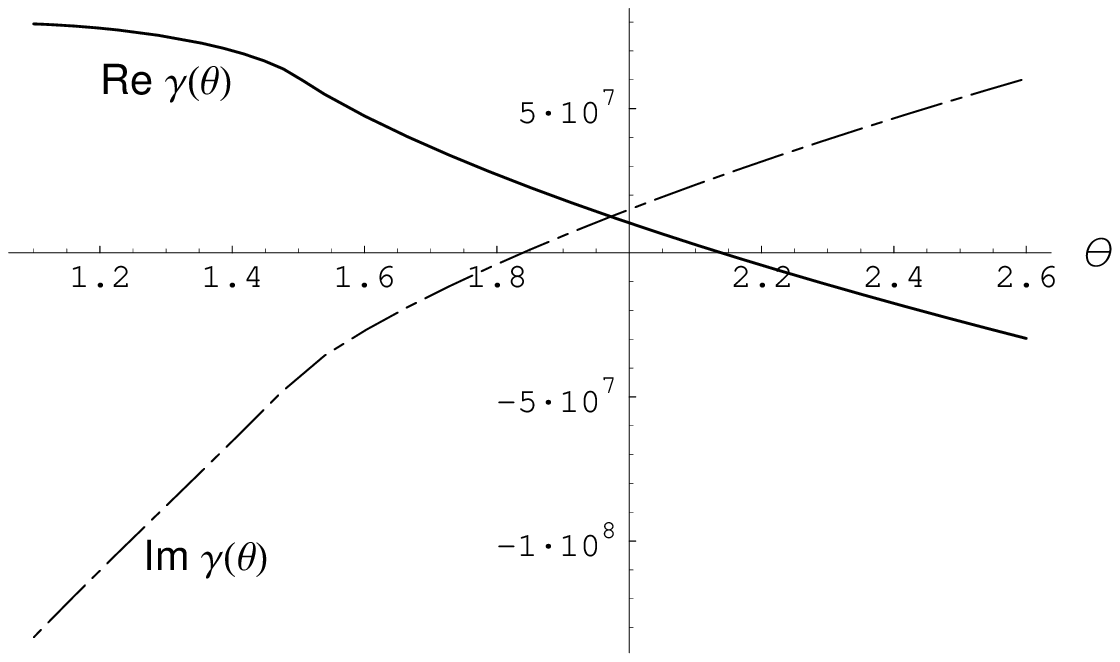}
\end{center}
\begin{center}\parbox{.9\textwidth}{Fig.1 \textit{Real and imaginary parts of the function
$\gamma(\theta)$. Here, $\o_c=2.0\times 10^{16} s^{-1}$ and the
medium is described by Brillouin's choice of parameters.}}
\end{center}

\begin{center}
\includegraphics[width=.7\textwidth]{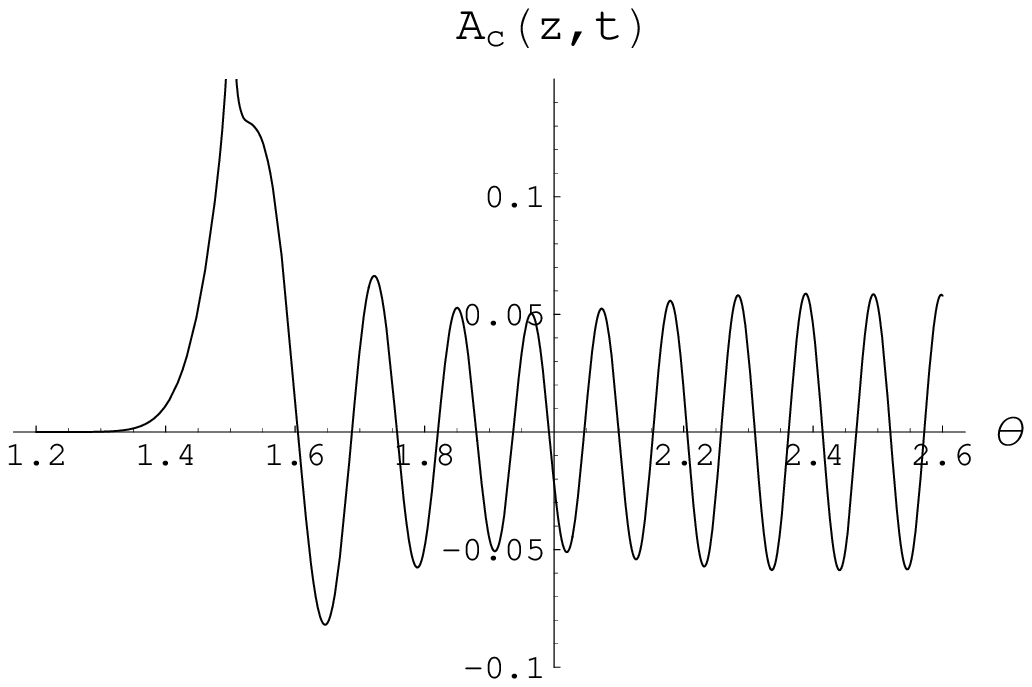}
\end{center}
\begin{center}\parbox{.9\textwidth}{Fig.2 \textit{Representation of the main signal
in the medium described by Brillouin's choice of parameters, based
on Eq.(33). Here, $\o_c=2.0\times 10^{16} s^{-1}$, $\l=3.0\times
10^{-15} s^{-1}$.}}
\end{center}
\end{figure}

\begin{figure}
\begin{center}
\includegraphics[width=.7\textwidth]{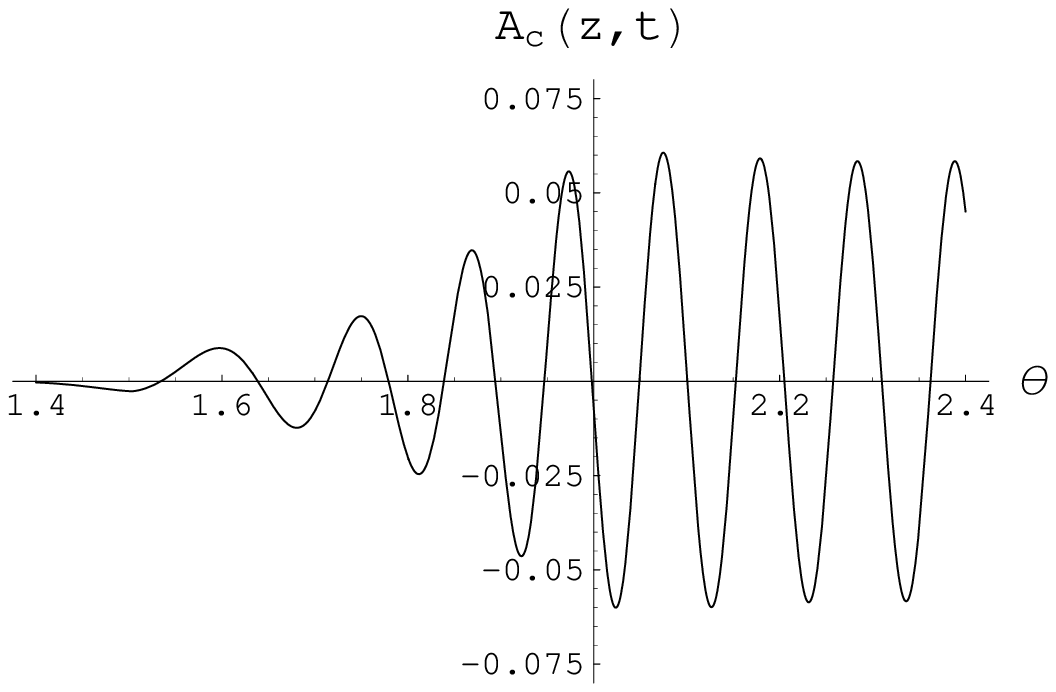}
\end{center}
\begin{center}\parbox{.9\textwidth}{Fig.3 \textit{Representation of the main signal
evolution in the medium described by Brillouin's choice of
parameters, based on Eq.(33) with deleted term proportional to
$(-\phi_{\o\o}(\os,\t))^{-1/2}$. Here, $\o_c=2.0\times 10^{16}
s^{-1}$, $\l=3.0\times 10^{-15} s^{-1}$.}}
\end{center}

\begin{center}
\includegraphics[width=.7\textwidth]{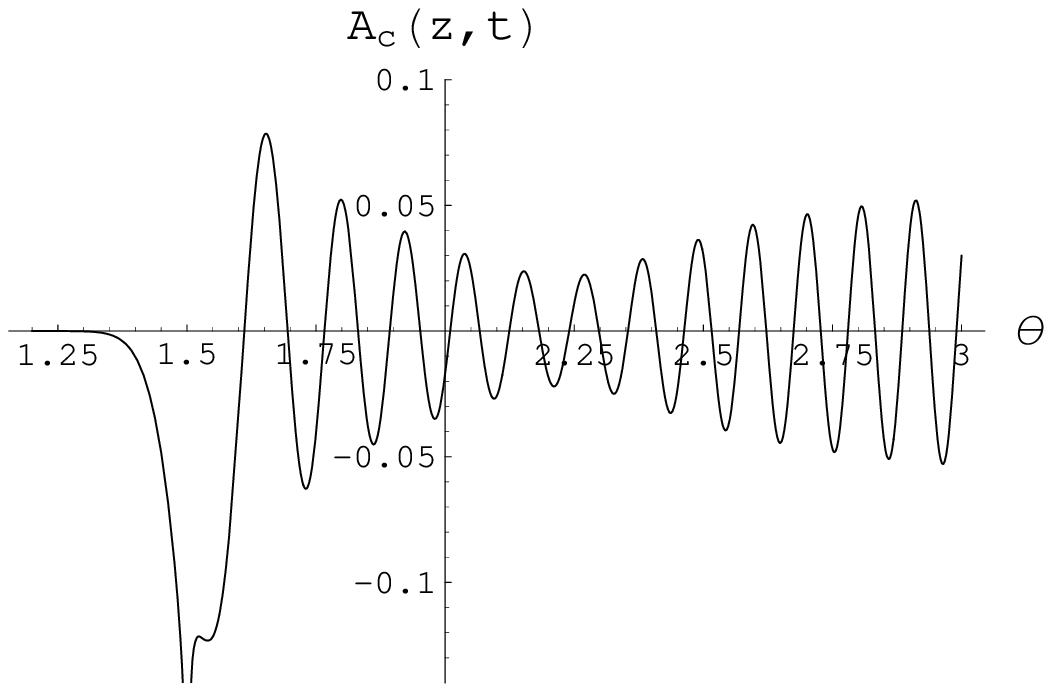}
\end{center}
\begin{center}\parbox{.9\textwidth}{Fig.4 \textit{Representation of the main signal
in the medium described by Brillouin's choice of parameters. Here,
$\o_c=2.0\times 10^{16} s^{-1}$, $\l=3.0\times 10^{-15} s^{-1}$
and $\beta=5.0\times 10^{14}$.}}
\end{center}
\end{figure}

Assume now that $\beta$ is sufficiently small, such that the
second pole at $\o_{c2}=\o_c-2i\beta$ can appear close to, or be
crossed by the deformed contour $P$. To fix our attention let
$\beta=5.0\times 10^{14} s$. In this case an expression similar to
(\ref{e34}) must be added to the asymptotic representation for
$A_c(z,t)$. By virtue of (\ref{e10}), the expression should be
multiplied by the factor $-2$, and $\o_{c2}$ should replace
$\o_c$.

The corresponding plot is shown in Fig.~4. It is seen that now the
growth of the main signal is slower then in Fig.~2.

\section{Conclusions}
In this paper the problem of electromagnetic signal propagation in
a dispersive Lorentz medium is considered. It is assumed that the
exciting signal is turned on at a finite time instant. The signal
rapidly oscillates and its envelope is described by a hyperbolic
tangent function. While propagating in the medium, the signal
splits into three components: Sommerfeld and Brillouin precursors,
and the main signal. In this work we find both non-uniform and
uniform asymptotic representations for the main signal evolution.
The former representation is readily obtainable by residues. The
latter representation is constructed with the help of
Bleistein-Handelsman method of uniform asymptotic evaluation of
integrals with nearby simple saddle point and an algebraic
singularity. We show, how the uniform representation, expressed in
terms of complementary error integral, reduces to the non-uniform
representation. The results here obtained are illustrated with a
numerical example. This paper is a complement to our earlier works
on Sommerfeld and Brillouin precursors (\cite{ac;2a},
\cite{ac;2b}).

\vspace{2ex}\noindent \textbf{Acknowledgment}

\noindent The research presented in this work was supported by the
State Committee for Scientific Research under grant 8 T11D 020 18.

\end{document}